\documentclass[conference]{IEEEtran}
\IEEEoverridecommandlockouts
\usepackage{cite}
\usepackage{amsmath,amssymb,amsfonts}
\usepackage{algorithmic}
\usepackage{graphicx}
\usepackage{textcomp}
\usepackage{xcolor}
\def\BibTeX{{\rm B\kern-.05em{\sc i\kern-.025em b}\kern-.08em
    T\kern-.1667em\lower.7ex\hbox{E}\kern-.125emX}}
\usepackage{geometry}
\begin{document}
\author{
\IEEEauthorblockN{1\textsuperscript{st} Yuanzhe Jin}
\IEEEauthorblockA{
\textit{Northwestern University}\\
Evanston, USA\\
yzjin@u.northwestern.edu}
\and
\IEEEauthorblockN{1\textsuperscript{st} Chenrui Zhang}
\IEEEauthorblockA{ 
\textit{Southeast University}\\
Nanjing, China\\
chenruizhang@seu.edu.cn}
\and
\IEEEauthorblockN{2\textsuperscript{nd} Maorong Wang}
\IEEEauthorblockA{ 
\textit{Northwestern University}\\
Evanston, USA\\
mrwang@u.northwestern.edu}
}
\title{Otaku: Intelligent Management System for Student-Intensive Dormitory}

\maketitle

\begin{abstract}
In most student dorms in developing countries, a large number of people live in single-function dorm units. The division of dormitory is too fixed, resulting in the dormitory often lacking functional spaces such as entertainment, sports, meetings, etc. At the same time, a large number of people are likely to cause aggregation at a fixed time, which is not conducive to maintaining social distance under pandemic conditions such as COVID-19. This brings a lot of inconvenience to students' life study and management staff.

In this paper, we present a smart dormitory system named Otaku using the Internet of Things technology to integrate facilities related to student dormitory life. By splitting the dormitory into several different categories according to their functionality by using smart door lock design, the system can achieve a more effective and flexible resource allocation, which not only helps the school management but also benefits students.
\end{abstract}

\begin{IEEEkeywords}
Internet of Things, Embedded Systems, Smart Dormitory, Intelligent Control, Smart Campus
\end{IEEEkeywords}

\section{Introduction}
In most developing countries and some developed countries, the “Student-Intensive” arrangement of college dormitory and related resources restricts students' freedom of study and life. The definition of the “Student-Intensive" dormitory is that the student's accommodation space is less than 5 square meters per person. This means that it is very difficult to implement social distance in the current situation of COVID-19.

In traditional dormitory buildings, all rooms are used as dormitory rooms and a small number of function rooms are arranged, which makes students relatively fixed in the dormitory. When students exchange dormitories or perform recreational activities, they show poor flexibility and waste of resources. However, since most school dormitory buildings have already been built, it is impractical to rebuild the dormitory buildings.

We need a way to renovate the dormitory building to the minimum to meet the students' diverse accommodation needs under the existing dormitory building conditions. If there are some space rooms, after deploying the whole system, each area of the dormitory area can be used as a dormitory "region". They can be regarded as an entertainment room, a study room, or a meeting room. Students can choose or trade freely, thereby achieving the purpose of reasonable allocation of dormitory resources. Managers can also benefit from it when happens unexpected incidents in the dormitory. A feasible method is to flexibly redistribute the originally fixed resources through the rapid development of smartphones and IoT hardware terminals.

\section{Related Work}
The smart dormitory is a system, using the Internet of Things technology to integrate facilities related to student dormitory life\cite{chen2012study} and build an efficient dormitory facility management system. It can improve the safety, convenience, and comfort of student accommodation, and achieve an energy-saving living environment\cite{agarwal2010occupancy}.

Researchers have researched the application of IoT devices on campus and found possible problems. Lee proposed a smart dormitory system\cite{lee2016implementation}. It pays more attention to the facilities in the dormitory, which means it does not just focus on the dormitory. Campus management and dormitory management have realities, but the difference is that the composition of dormitory management is more unitary and different from the complexity of school management\cite{yang2018situational}. 

Researchers also design applications of IoT technology to campus scenarios\cite{lin2012application}. It is helpful to use the application to keep track of public facilities on campus\cite{deling2014introduction}. IoT can not only help using public facilities but is also helpful as a studying tool for the student. Learners can enhance the context-aware learning process through the physical structure of the campus\cite{atif2013social}. Another structure uses a user-centering experimental research facility for IoT technology\cite{nati2013smartcampus}. It makes users have a stronger sense of substitution for the surrounding environment. The smart campus can even use social information. Some studies have described a flexible system architecture based on service-oriented specifications to support social interaction in a campus-wide environment\cite{6142288}. The university campus also has an interesting application environment for the ubiquitous computing paradigm, where a large number of users share a large number of information needs. The researchers put forward the idea of putting ubiquitous computing concepts into practical use and gain new insights into the design of virtual copies of real-world objects\cite{1203564}.

A smart dormitory can also cooperate with student education. With the development and application of cloud computing and the Internet of Things, some researchers have distinguished the difference between digital campus and smart campus. By building an intelligent campus model and application framework based on cloud computing and the Internet of Things, analyzing its functional applications to build a system based on a smart campus\cite{Nie2013/03}.

There are many similarities between smart dormitory management and smart office/building. And IoT technology can be helpful to reduce energy consumption\cite{6742619}. There is a special design architecture for demonstrating energy-saving policies on a building-wide basis through real-time energy prices and the location and preferences of each user\cite{chen2009design}. Other researchers present a systematic framework for a building energy monitoring and analysis system based on the Internet of Things\cite{6066758}. Some article provides a view of software systems that improve building energy efficiency. It proposes an architecture that allows for phased investment in technology to capture energy-saving benefits in various use cases\cite{5342106}. Besides, studies have shown that the possibility of applying underfloor air conditioning to tasks or personalized systems. This system can be used as a suitable choice for the air-conditioning system in future intelligent buildings\cite{osti_211849}.

The innovation of \textit{Otaku} lies in the application scenario. Today, the Internet of Things technology has been widely used in homes and offices\cite{buckman2014smart}. But in the area of the student dormitory especially in the room, there is still a lack of applications for applying IoT systems.

\section{system}
Otaku's design ideas apply to many IoT devices. The core of the design idea lies in decentralized management, which means that the client has independent rights between other clients and is not subject to a unified and centralized management center. Limited by research, we present the door lock designed with this idea in this article.

\subsection{System Architecture} 
\begin{figure}[htbp]
  \includegraphics[width=0.95\linewidth]{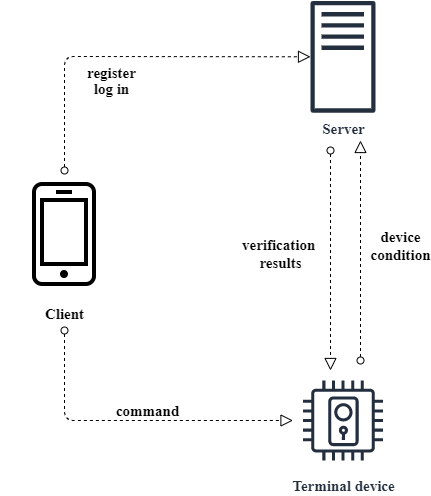}
  \caption{System Architecture}
  \label{fig:system architecture}
\end{figure}

The proposed system consists of three major parts: server, APP, and terminal IoT device. As is shown in Fig.\ref{fig:system architecture}.

The server stores the users’ and devices’ information, including username, pin code, users’ authority, devices’ name, devices’ occupancy condition, etc. The server can update users’ pin code and authority and send messages to corresponding devices.  

APP can communicate with the server to register an account, edit account information, send the request to apply authority to use the corresponding device, etc. It can also communicate with the terminal device to use them, send related commands. The administrator can use the APP to manage users' register requests and monitor terminal device work conditions, which helps daily maintenance.

Terminal IoT device acts as a hardware platform. It responds to users request, follows users’ command to work and send device condition, like if occupied, if work normally, to server. It can be attached to different sensors to be applied to different scenarios. For example, we can combine a mechanical lock with a relay and WiFi chip to build a smart dormitory lock; we can combine laundry machines with sensors and WiFi chips to build a smart laundry machine.

\subsection{Communications Protocol} 
WiFi is used to transmit data and command. WiFi is ubiquitous nowadays and especially in university, campus-wide WiFi is a common cause. A short-range and long-range combined communication mechanism is used for the system, which benefits from the abundant instruction set of ESP8266. Users can use different communication ways according to different scenarios.

For the server, The terminal itself stores a list of local equipment and related information, such as beds, controlled equipment, door locks, etc. in the dormitory. Therefore, the information that the server needs to send to the terminal is the users of various facilities in the dormitory and their personal information. Besides, when the server sends this information to the terminal, the terminal will store each facility as the basic unit, that is, each facility will have a white list that stores the user names and user rights that are authorized to use the item. Also, according to different user permissions, the degree of openness of the functions of the items is different, and some functions will only be open to administrators or users with special needs.

\begin{figure}[htbp]
  \includegraphics[width=0.45\linewidth]{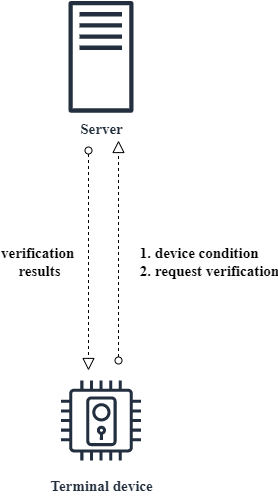}
  \centering
  \caption{A server to ternimal scheme}
  \label{server}
\end{figure}

For the terminal, the control request information for the terminal is sent directly from the client to the terminal without passing through the server, so all the client's operations on the terminal and the status of the terminal's various items must be sent to the server in real-time. The process of terminal and server is shown in Fig.\ref{server}.

For the client, the information communicated between the client and the terminal is relatively single. Generally, the client sends control information to the terminal in one direction. It mainly includes user name and control request information. The terminal stores a local user white-list. The client also sends its user name by default when sending control information. Therefore, when the client sends control information to the terminal item, the terminal verifies whether the user is on the white-list. If it is, the function within the user's authority is executed against the control information sent by the user; if not, the failure information is fed back to the client.

\begin{figure}[htbp]
  \includegraphics[width=0.8\linewidth]{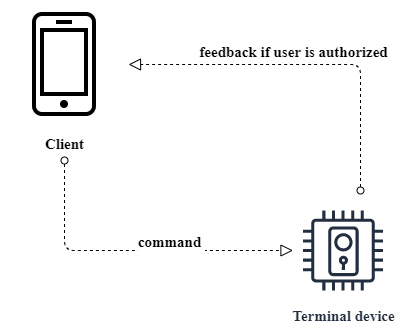}
  \centering
  \caption{A client to ternimal scheme}
  \label{client}
\end{figure}

The process of terminal and client communication to implement decentralized management and user authentication is shown in the Fig.\ref{client}.

\subsection{Management Mechanism } 
The key idea of the system is the decentralization management mechanism. In the design of Otaku, all facilities are equipped with sensors and WiFi chip, which are so-called IoT terminal devices. All these terminal devices are managed separately. For every device, users can apply the authority to use it if it is available and not occupied. 

\subsection{Communication Mechanism}
The management system uses WiFi to transmit data and command. WiFi is ubiquitous nowadays and especially in university. The short-range and long-range combined communication mechanism for the system can benefit from the abundant instruction set of ESP8266. Users can use different communication ways according to different scenarios.

The short-range communication mode is based on the ESP8266 AP mode communication mode. In this case, ESP8266 acts as a router to create a WiFi hot spot network that can be accessed by other STATION devices to form a local area network. The long-range communication mode is based on the ESP8266 STATION mode. At this time, ESP8266 acts as a STATION connected to the router as the AP, and maps ESP8266 to the external network through the Intranet penetration, to achieve the function of remote control.

There are two ways for intranet penetration methods. One is to set the router as a DMZ host. After finding out the IP of the ESP8266 in the internal network, set the router, perform DMZ forwarding, and enter the IP of the ESP8266 just queried in the DMZ host settings. After completing the settings, the user can access the internal network from the external network and complete remote control. The DMZ host builds a bridge between the external network and the internal network server. It is an internal network buffer that can be accessed by the external network while protecting the internal network.

However, this solution has some drawbacks. The router's WAN IP must be a public network IP. The earliest public networks were public IPs, but with the popularity of computers, the operators did not have enough public IPV4 addresses assigned to users. They could only privately change public IPs to intranet IPs and use them for multiple users.  Secondly, after the router is powered off and powered on again, the public network IP may change. Therefore, if intranet penetration is to be achieved, the IP address needs to change at any time. 

The other solution is to use Peanut Shell's intranet penetration and domain name resolution services. Using Peanut Shell to set up intranet penetration is similar to setting up DMZ forwarding on the router. The principle is the same, so only the domain name resolution service of Peanut Shell is introduced here. When using the Internet, humans use human language to remember the network, that is, the domain name, but communication between machines can only be done through machine language, which is the number-IP address. A domain name is just an external package, which is a webspace IP inside, and domain name resolution is a service that interprets human language for machines so that humans can use domain names and machines to use IP addresses to complete the conversion. Domain name resolution is done automatically by a dedicated server. The use of Peanut Shells is also relatively simple. The domain name resolution setting is on the same interface as the intranet penetration setting. Just enter the available domain name under the user name, the user can map the IP of the internal network to the external network through the router, and have a fixed domain name. This avoids the disadvantages caused by IP changes after the router is powered off and restarted. So after comparing the two solutions, this paper chooses to use Peanut Shell's intranet penetration as the method.

\subsection{System Use}
In the dormitory scenario described in this research, even in the same dormitory, many areas are separate. Different areas or controlled devices have their ownership and control rights, and at the same time, roommates all have permission to open and close the door lock. Therefore, the idea of decentralized use in this solution is different from the traditional IoT design concepts such as smart homes or smart hotels. Unlocking the door is one of the basic applications of decentralized management. 

\begin{figure}[htbp]
  \includegraphics[width=\linewidth]{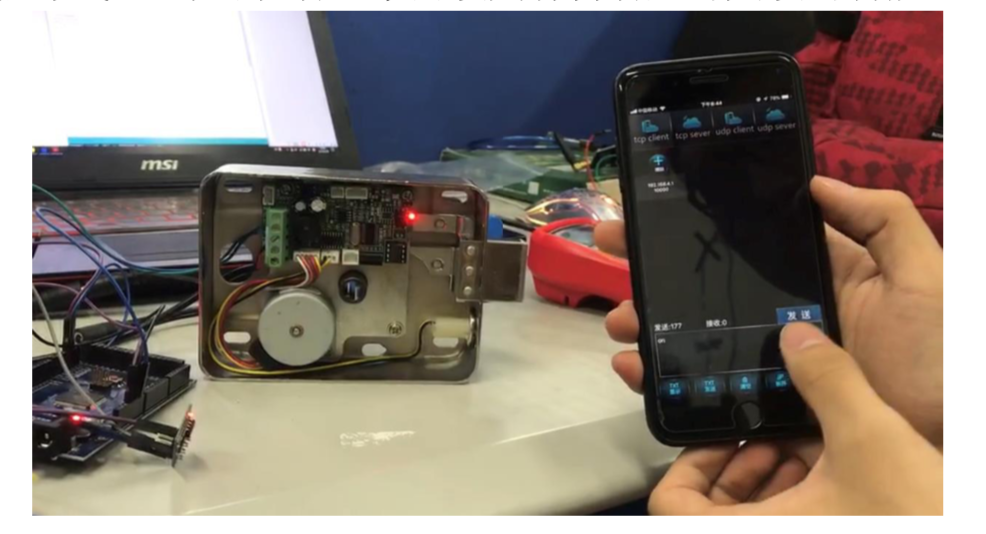}
  \centering
  \caption{Students use the mobile phone to control the door lock}
  \label{doorlock}
\end{figure}

The authority setting of the traditional IoT solution is often limited to the switch of the door lock. A flexible and efficient gradient permission management mechanism is not only compatible with the basic user requirements of the IoT platform in the dormitory scenario, but also one of the core to promote the rapid operation of the overall system. For example, if a student wants to open the door of their dormitory, log in to the client, and send an unlock request to the terminal. After the terminal verifies the student ’s user name and permissions through the whitelist that has been stored locally, the terminal successfully unlocks and locks the user Name (Joe), request information content (unlock), operation result (success), which is shown in Fig.\ref{doorlock}.

\section{Evaluation of the system}
A user study is shown in Fig.\ref{fig:User Study} about the user's experience of the dormitory management system.

\subsection{User Study}
\begin{figure}[htbp]
  \includegraphics[width=\linewidth]{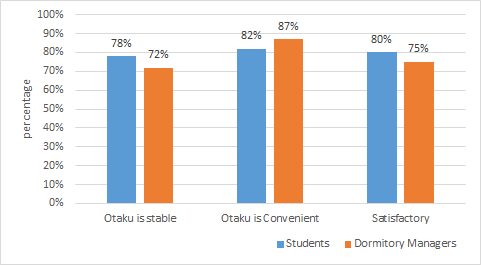}
  \caption{The result of the user study}
  \label{fig:User Study}
\end{figure}

The system is evaluated by 29 participants from Southeast University, China. We had 21 undergraduate students (12 males and 9 females) and 8 dormitory managers live in a dorm equipped with Otaku for a whole week. The participants rated the system as convenient and stable, according to the results of the survey. The result of the survey is shown in Fig. \ref{fig:User Study}.

The 21 undergraduate participants are either juniors or seniors, with more than 2 years of dormitory living experience, and 8 managers are all skilled employees with more than 3 years of working experience as college dormitory managers.

As a result of the survey, over 78\% of students and 72\% of employees rate highly on the system's robustness and stability. Moreover, Nearly 85\% of students and 87\% of employees believe that this dormitory system has brought them some convenience. In addition to that, 80\% percent of undergraduate participants and 75\% of managers are extremely satisfied with the system and are willing to continue to use Otaku in their dormitory.

Due to the decentralization management method we implement, part of users' information will be stored in the terminal device. Hence, potential risks about the terminal device being hacked exist. Besides, the system is based on robust and dormitory-covered WiFi support, which means in some traditional dormitories they may not supply WiFi service or in some situations that the electricity is not stable, so the WiFi signal can not hold. If not work under a stable WiFi condition, the terminal device will be in limited use. In response to some sudden power failure, students can turn back to use it as a traditional lock. 

In brief, the user study investigated the acceptability of the system, and most of the participants rate highly on the systems' ability to bring convenience. This demonstrates that this system has potential use for most schools.

\subsection{System Cost Evaluation}
The reason why large-scale intelligent dormitories cannot be promoted before is because of excessive cost. They are not suitable for student dormitories and schools where they cannot afford to purchase an extra dormitory management system. To solve the cost problem, we use some existing devices by attaching sensors into a network device instead of replacing all old devices into IoT products. 

The terminal device is built by MCU, WiFi chip, and related sensors or mechanical components. The hardware platform is mainly composed of the Arduino Mega2560 development board and the ESP8266WiFi module. Since the system is built on the existing campus wireless network system, the main cost of the system lies on the terminal device. The average cost of the system is low. 

A single smart dormitory management system consists of two parts, an Arduino Mega 2560 develop board and an ESP 8266 WiFi chip. Both components are of low cost and sufficient supply. The average price of Arduino Mega 2560 is \$6 per chip, and the average price of ESP 8266 chip is \$2.24, which means it only takes \$8.24 for a dormitory room to have this management system equipped.

As a conclusion, this dormitory management system can be regarded as a cheap and affordable solution for most college dormitories.
\section{conclusion}
In the experiment and experience, it can be found that this intelligent management system is user-friendly with a cheap cost. Otaku enables convenient dormitory management and is proved to be convenient and satisfactory. It can be seen that the deployment of this decentralized management idea on the door lock is feasible. It would be potentially helpful to apply this idea to other public equipment such as washing machines, game consoles, and cooking utensils. 

In future improvement, it is helpful to take advantage of the occupancy information that Otaku collects and design some energy-saving logic. For example, the HVAC system and the light system based on the location information of users are useful to achieve power saving. With the system being deployed, the student dormitory can achieve more flexible and effective resource allocation.

\nocite{E.2007}
\nocite{}
\bibliographystyle{unsrt}
\bibliography{refs}

\end{document}